\documentclass[twocolumn]{aastex62}

\newcommand{\GALEX}{{\it GALEX}}
\newcommand{\Gaia}{{\it Gaia}}

\newcommand{\HST}{{\it HST}}

\newcommand{\uBVI}{{\it uBVI}}

\shorttitle{The ZNG Catalog of UV-Bright Stars}
\shortauthors{Bond}

\begin{document}          

\title{Astrometric Membership Tests for the Zinn--Newell--Gibson \\
``UV-Bright'' Stars in Galactic Globular Clusters}

\correspondingauthor{Howard E. Bond}
\email{heb11@psu.edu}

\author[0000-0003-1377-7145]{Howard E. Bond}  
\affil{Department of Astronomy \& Astrophysics, Pennsylvania State
University, University Park, PA 16802, USA}
\affil{Space Telescope Science Institute, 3700 San Martin Drive,
Baltimore, MD 21218, USA}
\affil{Visiting Astronomer, Cerro Tololo Inter-American Observatory and Kitt Peak National Observatory, National Optical Astronomy Observatory, operated by the Association of Universities for Research in Astronomy under a cooperative agreement with the National Science Foundation.}

\begin{abstract}

In 1972, Zinn, Newell, \& Gibson (ZNG) published a list of 156 candidate ``UV-bright'' stars they had found in 27 Galactic globular clusters (GCs), based on photographs in the $U$ and $V$ bands. UV-bright stars lie above the horizontal branch (HB) and blueward of the asymptotic-giant branch (AGB) and red-giant branch in the clusters' color-magnitude diagrams. They are in rapid evolutionary phases---if they are members and not unrelated bright foreground stars. The ZNG list has inspired numerous follow-up studies, aimed at understanding late stages of stellar evolution. However, the ZNG candidates were presented only in finding charts, and celestial coordinates were not given. Using my own collection of CCD frames in $u$ and $V$, I have identified all of the ZNG objects, and have assembled their coordinates, parallaxes, and proper motions from the recent \Gaia\/ Early Data Release~3 (EDR3). Based on the \Gaia\/ astrometry, I have determined which objects are probable cluster members (45\% of the sample). For the members, using photometry from EDR3, I have assigned the stars to various evolutionary stages, including luminous post-AGB stars, and stars above the HB\null. I point out several ZNG stars of special interest that have still, to my knowledge, never been studied in detail. This study is an adjunct to a forthcoming survey of the Galactic GCs in the \uBVI\/ photometric system, designed for detection of low-gravity stars with large Balmer discontinuities.

\end{abstract}

\keywords{stars: AGB and post-AGB --- globular clusters --- stars: evolution --- post-horizontal-branch stellar evolution  }

\section{Stars Above the Horizontal Branch in Globular Clusters}

In color-magnitude diagrams (CMDs) of globular clusters (GCs), the vast majority
of stars in post-main-sequence evolutionary stages lie on the subgiant branch,
red-giant branch (RGB), horizontal branch (HB), and asymptotic-giant branch
(AGB)\null. There are, however, rare stars in transient phases of rapid evolution that lie
above the HB and to the blue of the AGB\slash RGB in cluster CMDs. These objects include stars that have departed the HB and are evolving toward the AGB, post-HB stars that
reached the AGB but then left it before arriving at the AGB tip (post-early-AGB,
or PEAGB, stars), and objects that reached the tip of the AGB and are now
evolving rapidly toward higher temperatures (post-AGB, or PAGB, stars). See
\citet[][hereafter M+19; their Figure~7]{Moehler2019} for examples of theoretical post-HB evolutionary tracks that produce stars above the HB in GC CMDs. It is also possible that binary interactions can generate stars lying above the HB.

Our group at Pennsylvania State University is conducting an observational survey
aimed at creating a complete census of above-horizontal-branch (AHB) stars in the
Galactic GC system. This survey is based on photometry in the \uBVI\/ system
\citep{Bond2005}, which is optimized for detection of low-gravity evolved stars with
large Balmer discontinuities in their spectral-energy distributions. With astrometry from
\Gaia\/ now available, including the recent Early Data Release~3 (EDR3; \citealt{Gaia2020}),
we can further test cluster membership using precise parallaxes and proper motions (PMs). Our \uBVI\/ work led to
discoveries of luminous ``yellow'' PAGB stars in M79 \citep[][hereafter
BCS16]{Bond2016} and M19 \citep[][hereafter B+21]{Bond2021}. In the case of M19, we also
discovered a hotter, and equally luminous, blue PAGB star belonging to the
cluster. Full details of our \uBVI\/ survey for AHB stars will be presented in a
forthcoming paper \citep[][hereafter D+21]{Davis2021}. Subsequent publications
will explore the utility of yellow PAGB stars---which are the visually brightest
stars in old populations---as standard candles for determining extragalactic
distances, as I have advocated \citep{Bond1997a,Bond1997b}.

\section{The Zinn-Newell-Gibson ``UV-Bright'' Stars}

The two luminous PAGB stars that B+21 found in M19 are among the brightest and most conspicuous
members known in any GC\null. In spite of this, they had not, to our knowledge, been
recognized as cluster members by previous investigators of this massive GC\null. However,
when we researched the literature, we found that both of the M19 objects had
been identified as ``UV-bright'' candidates in the classical survey of \citet[][hereafter ZNG]{Zinn1972}.

The ZNG team had blinked photographs of 27 GCs obtained in the $U$ and $V$
bands, and had identified 156 candidates that were the brightest objects in
the clusters in the $U$ band. The term ``UV-bright'' was a bit of a misnomer.
Many of the ZNG objects are indeed hot post-AGB stars, including the
prototypical luminous blue stars Barnard~29 in M13, and von Zeipel 1128 in M3.
However, many other ZNG candidates are even cooler than the yellow Type~II Cepheids
in GCs, but nevertheless brighter than most of the cluster members in the $U$
band---including objects that are actually unrelated bright foreground stars.

ZNG presented finding charts for their candidates, but did not provide celestial
coordinates. There have been several follow-up studies of individual ZNG
objects (e.g., \citealt{Zinn1974}; \citealt{Harris1983}; \citealt{deBoer1985}; \citealt{Jasniewicz2004}; \citealt{Moehler2001,Moehler2010}; M+19;  and references therein), and several of these stars have proven to be of great astrophysical interest. These include, for example, several now well-known hot PAGB stars (see the lists in M+19). However, there has never, to my knowledge, been a
published list of the celestial coordinates of all of the ZNG stars.

As an adjunct to the D+21 study, I determined the coordinates of all of the ZNG
stars, and I present them here. I then used astrometry from \Gaia\/ EDR3 to test
the cluster memberships of the candidates. Lastly, I give an indication of the
evolutionary status of the stars that appear to be cluster members, based on
their locations in the CMDs, and I point out several unstudied objects of interest.

\section{Identifying the ZNG Stars}

I first identified each ZNG star in my \uBVI\/ CCD GC survey CCD images. {  These frames had been obtained} with 0.9-, 1.5-, and 4-m telescopes at Kitt Peak
National Observatory and Cerro Tololo Inter-American Observatory; they are
described in detail by B+21 and D+21. I blinked the $u$ and $V$ frames,
while comparing them with the ZNG finding charts. In nearly every case, the
stars were identified unambiguously; only a handful of objects near the cluster
centers were less certain. {  Approximate coordinates were then determined} using
images from the Space Telescope Science Institute Digitized Sky Survey,\footnote{\url{https://archive.stsci.edu/cgi-bin/dss_form}} images from the PanSTARRS-1 sky survey,\footnote{\url{https://ps1images.stsci.edu/}}
and/or in a few cases images from the {\it Hubble Space Telescope}.\footnote{\url{https://hla.stsci.edu/hlaview.html}} With all of these tools, coordinates can be obtained by placing a cursor on the stellar images.

Finally, I identified each object in the \Gaia\/ EDR3 catalog.\footnote{\url{https://vizier.u-strasbg.fr/viz-bin/VizieR-3?-source=I/350/gaiaedr3 }} Table~1 lists the 156 ZNG candidates, along with their J2000 coordinates, \Gaia\/ parallaxes and PMs, and the \Gaia\/ apparent $G$ magnitudes and $BP-RP$ colors. The notes to the table give information on the few instances of uncertain identifications, as well as other information on the objects, but are not intended to be a complete literature survey. 

\section{Cluster Membership Tests}

{ 
Using the EDR3 astrometry, I tested each ZNG candidate for cluster membership. Two  criteria were applied to each star: PM and parallax; and in a few instances a third one: radial velocity (RV)\null. First, I determined the mean PMs of each cluster in right ascension and declination and their dispersions, based on a selection of cluster members from the EDR3 catalog; then I required that each candidate have a PM consistent with that of the host cluster. In a large majority of cases, the PM criterion alone was sufficient to exclude the non-members. Next the parallax was considered. I adopted a nominal parallax for each cluster, based on the reciprocals of the distances given in the \citet[][hereafter H10]{Harris2010} catalog of cluster parameters.\footnote{Online version of 2010 December, at \url{http://physwww.mcmaster.ca/~harris/mwgc.dat}. {  Based on the discussion in \citet{GaiaHelmi2018}, I used the H10 cluster distances rather than distances based directly on \Gaia\/ parallaxes; in any case, this choice makes little difference in the membership tests.}} The parallax of each candidate was generally required to be consistent, within 3 times its stated
uncertainty, with the nominal cluster parallax. There was, however, a handful
of cases where the star's PM was in accord with membership, but the parallax
was larger (or occasionally smaller) than that of the cluster, by more than the
nominal 3 times the uncertainty. It is plausible that their EDR3 parallaxes may have been affected by source crowding. These ambiguous cases are discussed in the notes to Table~1. Finally, in a handful of instances for the brightest objects, EDR3 also gives a radial velocity (RV), another valuable membership criterion via comparison with the cluster's RV given in H10. The few cases where the \Gaia\/ RV helped confirm membership are indicated in the notes to Table~1.

Figure~1 illustrates the PM and parallax membership tests, for the case of the cluster M22. I selected stars from \Gaia\/ EDR3 lying within $5'$ of the cluster center, and brighter than $G$ magnitude 13. The PMs of these stars are plotted as black and red points: black for stars with a parallax less than 0.6~mas (which is about twice the nominal cluster parallax), and red for the few stars with a parallax of more than 0.6~mas (likely foreground stars). The EDR3 PMs are so precise that the cluster members form a tight distribution. The small number of field stars and foreground objects have a much wider dispersion and are offset from the cluster mean. 

The larger filled circles in Figure~1 plot the PMs of 16 of the 18 ZNG candidates. (Two of them have PMs so large they are outside the frame.) The ZNG stars plotted in blue have parallaxes less than 0.6~mas, and the ones with larger parallaxes are plotted in green. Four of the ZNG objects have PMs consistent with cluster memberships, as well as small parallax values, making them highly probable members. One of them, as noted in Table~1, also has a RV consistent with membership. The remaining ZNG candidates in the figure have PMs clearly inconsistent with that of the cluster, and most of them also have parallaxes too large for cluster membership.\footnote{There is always a small possibility of a field star happening to have a similar parallax and PM to that of the cluster, especially in cases of a cluster with a small PM superposed on a rich field; we will give a detailed and more formal discussion of cluster-membership probability estimation in D+21.}

\begin{figure}[hbt]
\begin{center}
\includegraphics[width=3.25in]{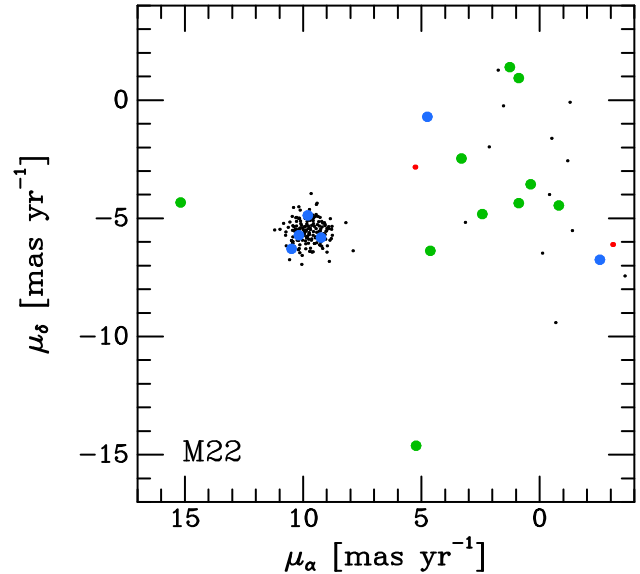}
\figcaption{\footnotesize       
\Gaia\/ EDR3 proper motions for stars lying within $5'$ of the center of the globular cluster M22. The small points plot stars brighter than $G=13$; black points are for stars with a parallax less than 0.6~mas, and red points for stars with larger parallaxes. The proper motions for 16 of the ZNG stars are plotted as large filled circles, colored blue for parallaxes less than 0.6~mas, and green for larger parallaxes. Two of the ZNG candidates have large proper motions placing them outside the frame. Four of the ZNG stars have both proper motions and parallaxes consistent with cluster membership.}
\end{center}
\end{figure}

The final membership classifications are given in column
10 of Table 1, with a question mark for the few ambiguous cases; these are
discussed in the notes to the table.

}

\section{Evolutionary Status}

For the ZNG candidates that I considered to be likely cluster members, I estimated
their evolutionary statuses based on their locations in the cluster CMDs. The
classification scheme that I adopted is based on cluster CMDs in which the
\Gaia\/ apparent magnitude $G$ is plotted against the color index $BP-RP$.
Figure~2 provides examples. Here the \Gaia\/ data are presented for three ZNG
GCs: M13, M53, and NGC\,5897. Also included is M79, which is not one of the ZNG clusters, but contains the luminous yellow PAGB star discovered by BCS16. All four clusters are only lightly reddened [$E(B-V)=0.02$, 0.02, 0.09, and 0.01, respectively, from H10]. For each cluster, I selected a
sample of likely members from the EDR3 catalog, with parallaxes and PMs consistent with membership.

\begin{figure*}[hbt]
\begin{center}
\includegraphics[width=3.25in]{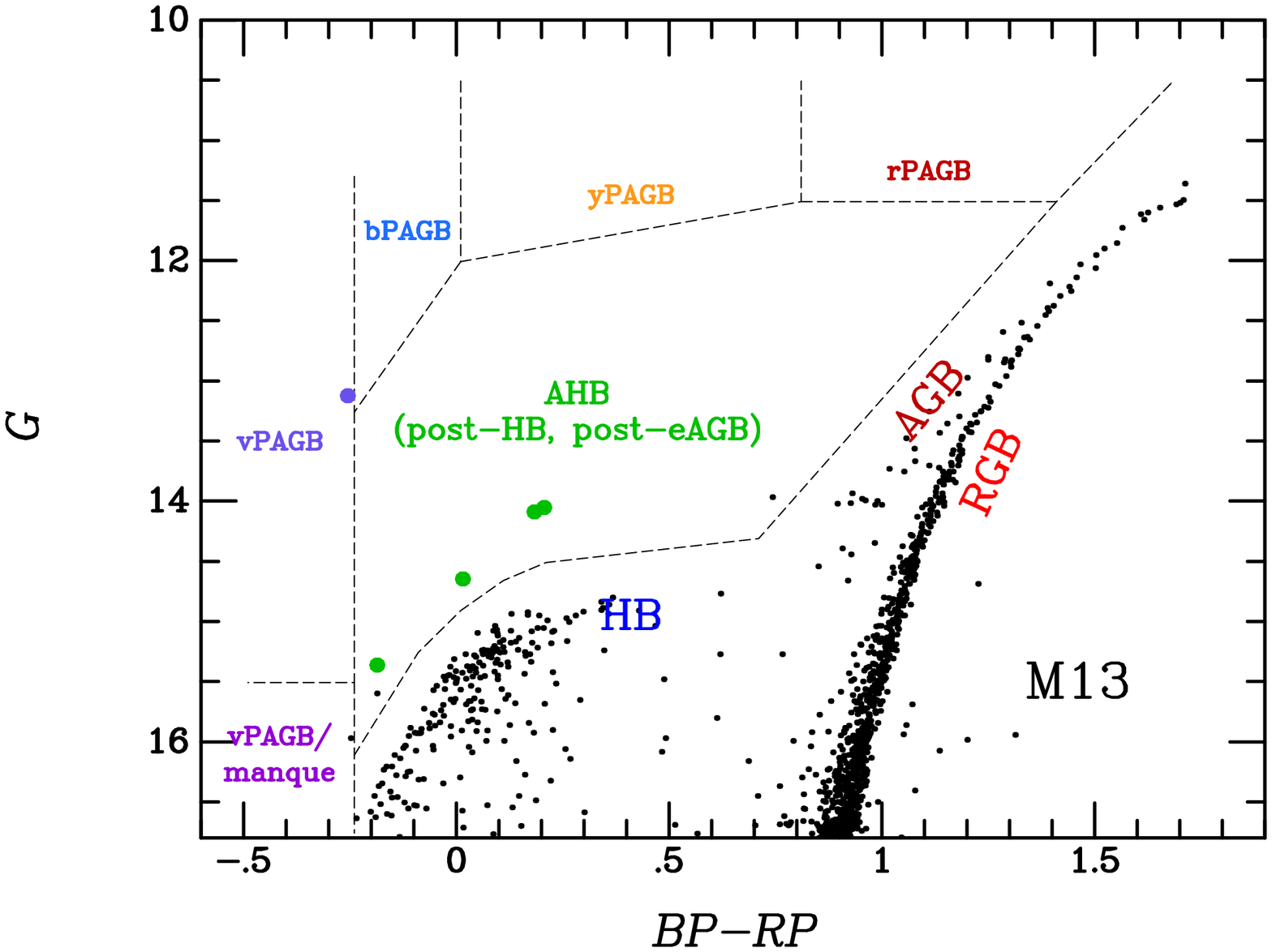}
\hfill
\includegraphics[width=3.25in]{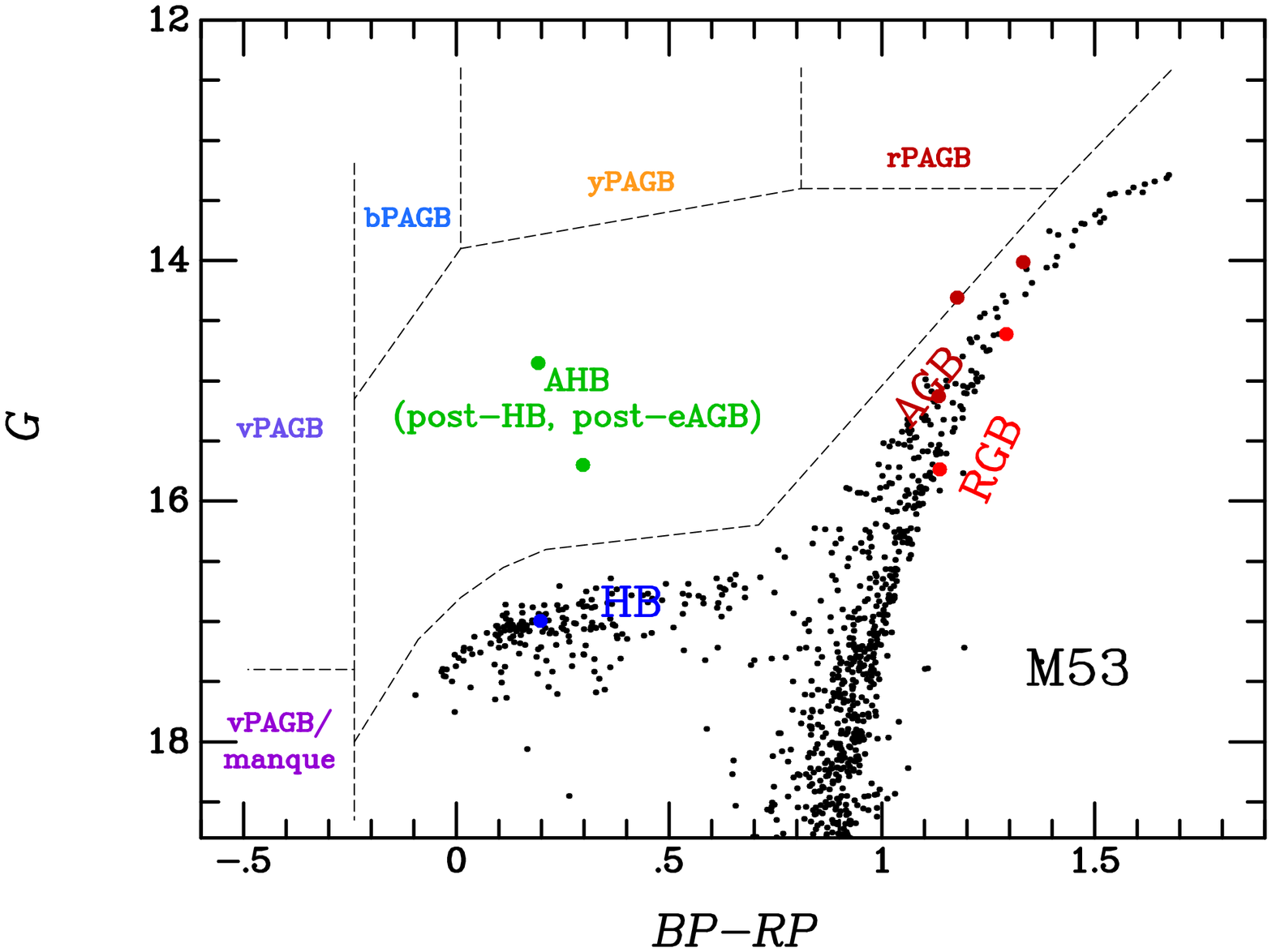}
\vskip0.2in
\includegraphics[width=3.25in]{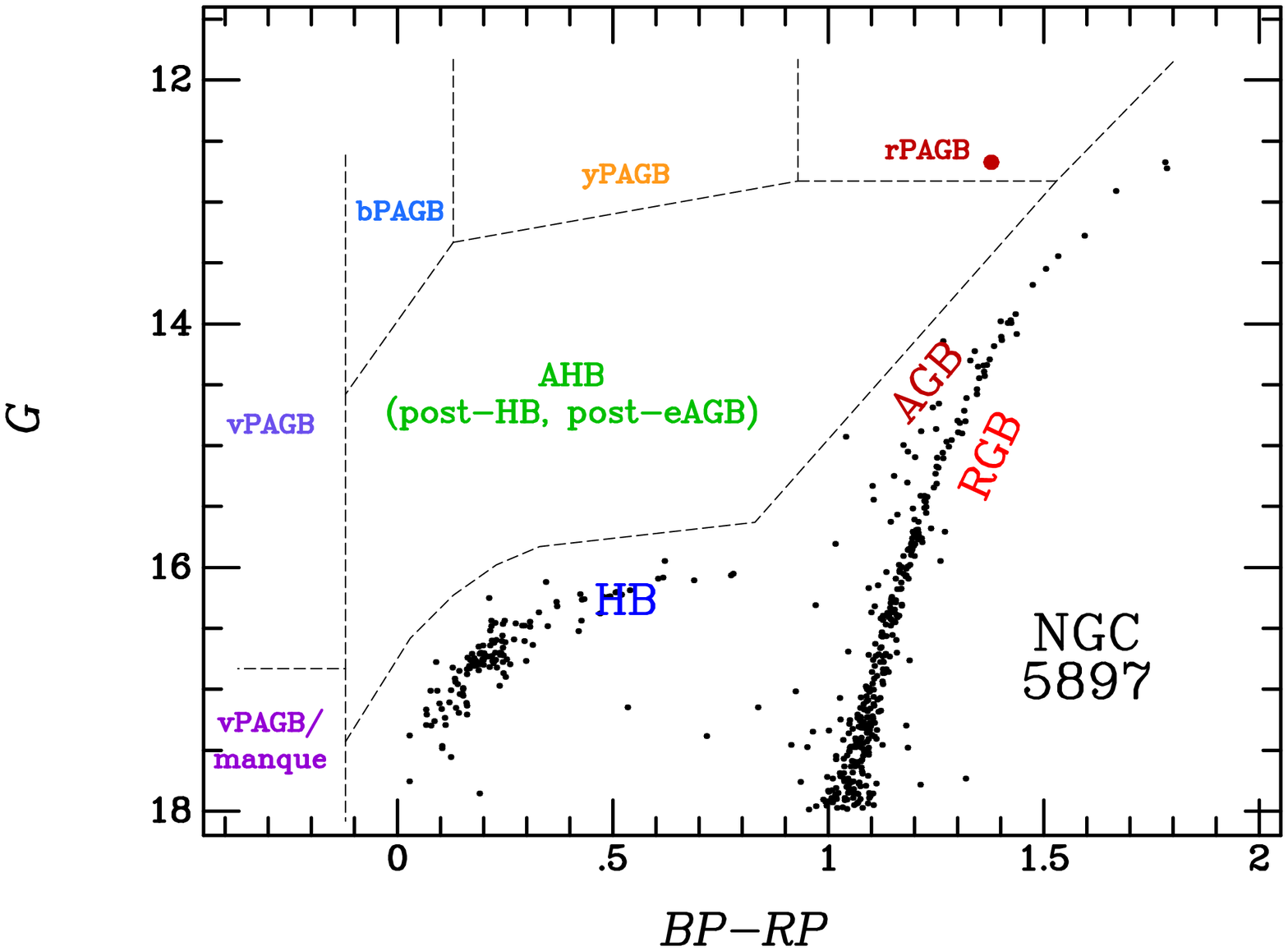}
\hfill
\includegraphics[width=3.25in]{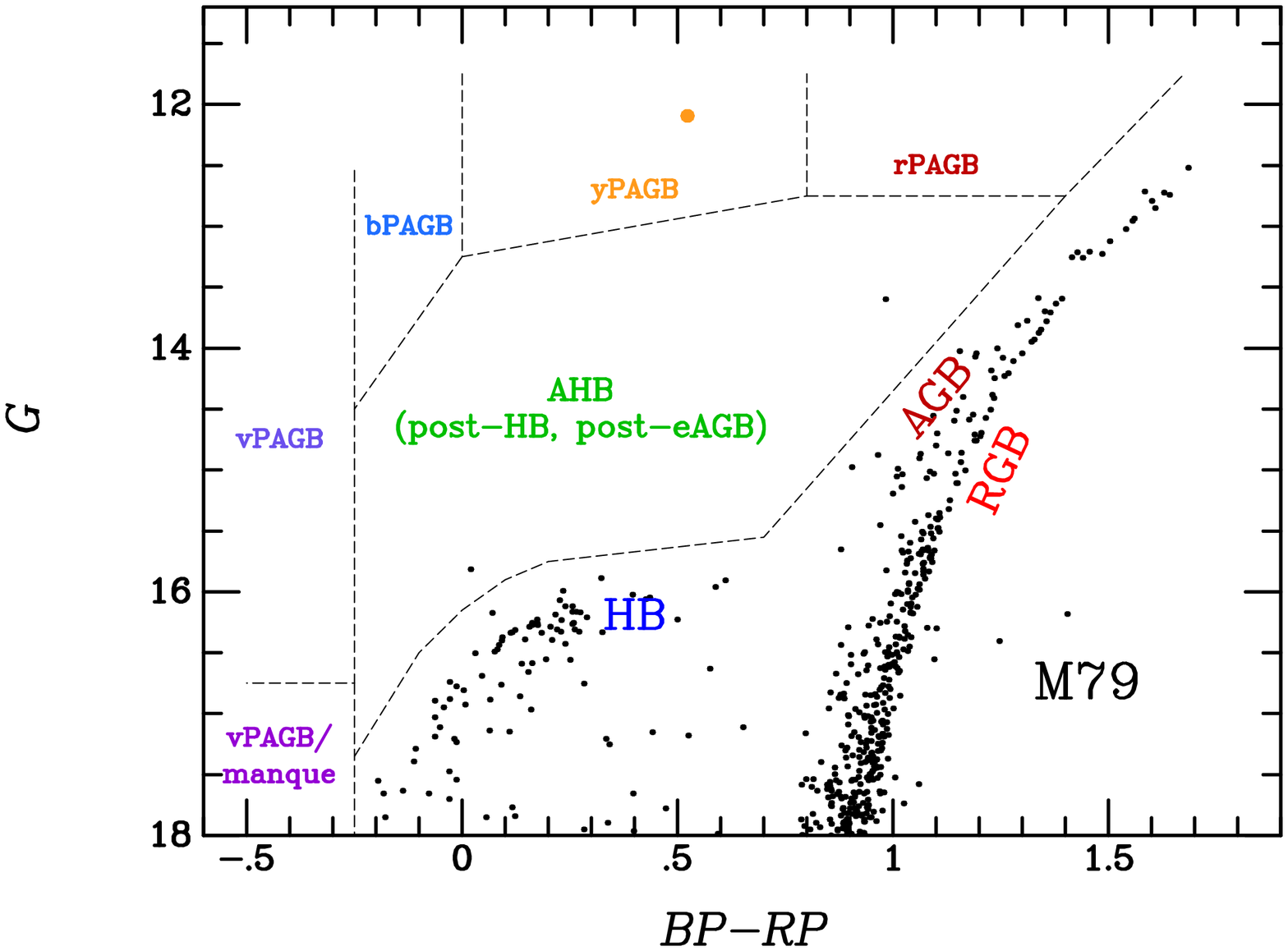}
\figcaption{\footnotesize
Examples of a schematic classification scheme for stars above the horizontal branch in globular clusters, based on color-magnitude diagrams plotting the \Gaia\/ $G$ magnitude versus $BP-RP$ color. The black filled circles in each frame show \Gaia\/ photometry for members of four lightly reddened clusters. The members were selected on the basis of \Gaia\/ EDR3 parallaxes and proper motions. The clusters' horizontal branch (HB), red-giant branch (RGB), and asymptotic giant branch (AGB) are labeled. The brightest stars are classified as post-AGB stars with red (rPAGB), yellow (yPAGB), blue (bPAGB), and violet (vPAGB) colors. Stars fainter than the PAGB sequence, down to within 0.5~mag of the HB or AGB, are classified as AHB objects; these are stars evolving off the HB toward cooler temperatures, along with stars leaving the AGB and evolving back to hotter temperatures. The hottest and optically faintest objects at the lower left are a combination of vPAGB stars and AGB-manqu\'e stars. See the text for further discussion of the scheme. The classification grid is shifted in color and magnitude according to the reddening and distance of each cluster. Upper left: M13. Filled colored circles mark the vPAGB star ZNG~1 = Barnard~29, and four AHB stars (ZNG~3, 4, 6, and 7). Upper right: M53. Filled colored circles mark two AHB stars (ZNG~3 and 4), two stars on the RGB (ZNG~6 and 9), three stars on the AGB (ZNG~7, 8, and 13), and one star on the HB (ZNG~16). Lower left: NGC\,5897. A filled colored circle marks the rPAGB star ZNG\,2. Lower right: M79. This cluster is not in the ZNG list, but is included to illustrate its luminous yellow PAGB star, discovered by BCS16.
}
\end{center}
\end{figure*}

In each panel in Figure~2, the region of the CMD above the HB and to the blue of the AGB is
divided schematically into several boxes. These are based on post-HB
evolutionary tracks, such as those in M+19. The luminous PAGB sequence is divided on the basis of color into
red, yellow, blue, and violet objects---rPAGB, yPAGB, bPAGB, and vPAGB,
respectively. The rPAGB stars have unreddened $BP-RP$ colors lying between 0.8 and up to within 0.1~mag of the AGB, and $G$ magnitudes at least 3.5~mag brighter than that of the HB at $BP-RP=0.8$. The yPAGB stars have $0<(BP-RP)<0.8$, and lie at least 3.5~mag brighter than the HB at the same color. bPAGB objects are in the color range $-0.25<(BP-RP)<0$, and are at least 3.5~mag brighter than the HB at the same color. The hottest PAGB stars are those bluer than $BP-RP=-0.25$, and either brighter than the brightness of the HB at a color of zero (vPAGB), or fainter (vPAGB/AGB manqu\'e). AGB-manqu\'e objects evolve directly from the extremely blue HB to the white-dwarf (WD) sequence, without ever becoming cool stars. As it happens, however, the ZNG sample does not contain any such stars. These divisions are somewhat arbitrary, but give a useful indication of the approximate temperatures. The scheme should be considered preliminary, and will be revised in our forthcoming paper (D+21).


Stars lying below the PAGB sequence, but at least 0.5~mag above the HB or AGB, are
called AHB objects. They are a mixture of stars evolving off of the blue HB to
cooler temperatures, along with PEAGB stars returning from the AGB toward the top of the
WD cooling sequence. 

\section{Results \label{sec:results}}

Using the plotting tools at Vizier (see footnote 4) to generate CMDs derived from \Gaia\/ EDR3
data for each GC, I applied the classification scheme described above to all of the ZNG
objects. The classification grid was shifted in color and magnitude according to the reddening and distance of each cluster. The classification results are given in column~11 of Table~1.

Out of the 156 candidates, 86 (55\%) are non-members or likely non-members, nearly all of them bright stars that clearly lie in the cluster foregrounds. Although ZNG excluded stars known to be variable at the time of their survey, the list contains several objects subsequently found to vary. These include two Type~II Cepheids and, more surprisingly, ten RR~Lyrae variables. I speculate that the latter were selected by ZNG because they happened to be near maximum light at the epoch of the
$U$-band exposures.

There are several cases of objects that are blends of two or more stars, which would account for them appearing unusually bright in ground-based images. More surprising is that there are quite a few candidates that appear to be unexceptional stars lying on the RGB, AGB, or HB sequences. However, a few of them are of interest because they are extremely bright,
lying close to the tips of the RGB or AGB\null. One of the most remarkable of these is NGC\,6712 ZNG~4, a very luminous star at the cluster's AGB tip; it is the known Mira variable AP~Sct, which is NGC\,6712 V2 in the \citet{Clement2001} catalog\footnote{Updated version available online at
\url{http://www.astro.utoronto.ca/~cclement/cat/listngc.html}} of variable stars
in GCs. One blue object, NGC\,6093 (M80) ZNG\,3, lies
close to the position of a known X-ray source, CX~3. 

Among the bright PAGB stars, there are two red PAGB (rPAGB) objects: NGC~5897
ZNG~2 and NGC~6626 (M28) ZNG~5. The former star has not, to my knowledge, been
recognized previously as a luminous PAGB star; the latter is a known RV~Tauri
variable, as described in the notes to Table~1. There is only one yellow PAGB
(yPAGB) star, the one in the cluster M19 that was recently pointed out for the first time by B+21. 

The table contains six
luminous blue PAGB (bPAGB) and two violet PAGB (vPAGB) stars. Of these, several
are already well known (see M+19): NGC\,5272 (M3) ZNG~1 (von Zeipel 1128),
NGC\,6205 (M13) ZNG~1 (Barnard~29), NGC\,6254 (M10) ZNG~1, NGC\,6712 ZNG~1, and
NGC\,6779 (M56) ZNG\,2. The bPAGB star NGC\,6273 (M19) ZNG~2 was only noted for
the first time in the recent B+21 study. NGC\,6402 (M14) ZNG~1 was identified as
a hot PAGB star based on \GALEX\/ imaging by \citet{Schiavon2012}, but has not
otherwise been studied. The bPAGB object NGC\,6333 (M9) ZNG~1 has not, to my
knowledge, been recognized previously. 

Lastly, Table~1 contains 27 AHB objects. As noted above, these are likely to be a mixture of stars evolving off the hot HB toward the AGB, and stars that reached the AGB and are now evolving back toward higher effective temperatures. Like the PAGB stars, they are worth further study.

The ZNG survey was an important early observational investigation that added to our understanding of late stellar evolution. Even after nearly five decades, there are still unexplored objects of considerable interest in the ZNG catalog. Our group's \uBVI\/ survey, to be reported by B+21, will follow in the path illuminated by this classical investigation.

\acknowledgments

I thank the Penn State ``post-AGB group''--- Robin Ciardullo, Brian Davis, and Michael Siegel---for discussions of evolved stars in globular clusters.

My \uBVI\/ observations were partially supported by NASA grant NAG 5-6821 under the ``UV, Visible, and Gravitational Astrophysics Research and Analysis'' program, and by the Director's Discretionary Research Fund at STScI\null. I also thank the staffs at Cerro Tololo and Kitt Peak for their support over many years.

This research has made use of the VizieR catalogue access tool, CDS, Strasbourg, France (DOI: 10.26093/cds/vizier). The original description of the VizieR service was published {  by \citet{Ochsenbein2000}.}

This work has made use of data from the European Space Agency (ESA) mission
{\it Gaia\/} (\url{https://www.cosmos.esa.int/gaia}), processed by the {\it Gaia\/}
Data Processing and Analysis Consortium (DPAC,
\url{https://www.cosmos.esa.int/web/gaia/dpac/consortium}). Funding for the DPAC
has been provided by national institutions, in particular the institutions
participating in the {\it Gaia\/} Multilateral Agreement.

The Digitized Sky Surveys were produced at the Space Telescope Science Institute under U.S. Government grant NAG W-2166. The images of these surveys are based on photographic data obtained using the Oschin Schmidt Telescope on Palomar Mountain and the UK Schmidt Telescope. The plates were processed into the present compressed digital form with the permission of these institutions.

The Pan-STARRS1 Surveys (PS1) and the PS1 public science archive have been made possible through contributions by the Institute for Astronomy, the University of Hawaii, the Pan-STARRS Project Office, the Max-Planck Society and its participating institutes, the Max Planck Institute for Astronomy, Heidelberg and the Max Planck Institute for Extraterrestrial Physics, Garching, The Johns Hopkins University, Durham University, the University of Edinburgh, the Queen's University Belfast, the Harvard-Smithsonian Center for Astrophysics, the Las Cumbres Observatory Global Telescope Network Incorporated, the National Central University of Taiwan, the Space Telescope Science Institute, the National Aeronautics and Space Administration under Grant No. NNX08AR22G issued through the Planetary Science Division of the NASA Science Mission Directorate, the National Science Foundation Grant No. AST-1238877, the University of Maryland, Eotvos Lorand University (ELTE), the Los Alamos National Laboratory, and the Gordon and Betty Moore Foundation.

Based in part on observations made with the NASA\slash ESA {\it Hubble Space Telescope}, obtained from the data archive at the Space Telescope Science Institute. STScI is operated by the Association of Universities for Research in Astronomy, Inc.\ under NASA contract NAS 5-26555.

\facilities{CTIO 0.9m, 1.5m; KPNO 0.9m, 4m;  Gaia, Swift, HST }


\def\p{\pm}

%

\newcount\remarkn
\newcount\noten


\def\countremarks{\remarkn=-53}

\def\rn{\global\advance \remarkn by 1 \number\remarkn}

\def\countnotes{\noten=0}

\def\nn{\global\advance \noten by 1 \number\noten}

\countremarks
\countnotes

\startlongtable

\begin{deluxetable*}{lccccccccccc}
\tablewidth{0 pt}
\tabletypesize{\tiny}
\tablecaption{The ZNG UV-Bright Candidates and \Gaia\/ Astrometry and Photometry
\label{table:zng}
}
\tablehead{
\colhead{Cluster} & 
\colhead{ZNG} &
\colhead{RA} &
\colhead{Dec} & 
\colhead{Parallax} &
\colhead{$\mu_\alpha$} &
\colhead{$\mu_\delta$} &
\colhead{$G$} &
\colhead{$BP-RP$} &
\colhead{Cluster} &
\colhead{Classi-} &
\colhead{Remarks\tablenotemark{c}} \\
\colhead{NGC/M} & 
\colhead{No.} &
\colhead{[J2000]} &
\colhead{[J2000]} & 
\colhead{[mas]} &
\colhead{[$\rm mas\,yr^{-1}$]} &
\colhead{[$\rm mas\,yr^{-1}$]} &
\colhead{[mag]} &
\colhead{[mag]} &
\colhead{Member?\tablenotemark{a}} &
\colhead{fication\tablenotemark{b}} &
\colhead{} 
}
\startdata
2419&1 &07 38 11.156&$+38$ 52 27.74&$ 0.5929\p0.0633$&$  4.740\p0.057$&$ -8.559\p0.036$& 16.318&$ 0.987$&no   &       &      \\ 
    &2 &07 38 09.234&$+38$ 53 12.77&$ 0.2301\p0.1527$&$ -0.257\p0.135$&$ -0.414\p0.090$& 17.395&$ 0.305$&yes  & AHB          \\
\noalign{\smallskip}
4147&1 &12 09 58.479&$+18$ 34 08.78&$ 1.0151\p0.0248$&$-20.213\p0.026$&$ -8.615\p0.023$& 14.326&$ 0.823$&no   &              \\
    &2 &12 10 05.661&$+18$ 32 45.69&$ 0.1253\p0.1069$&$ -1.858\p0.105$&$ -2.430\p0.099$& 16.856&$ 0.250$&yes  & HB    &  \rn \\
\noalign{\smallskip}
4590/&1 &12 39 40.099&$-26$ 43 02.62&$ 0.3911\p0.0289$&$-11.104\p0.030$&$ -5.431\p0.021$& 14.300&$ 0.498$&no   &              \\
M68 &2 &12 39 22.513&$-26$ 45 11.87&$ 0.0916\p0.0160$&$ -2.792\p0.016$&$  1.943\p0.012$& 12.258&$ 1.451$&yes  & AGB   &  \rn \\
\noalign{\smallskip} 
5024/&1 &13 12 56.879&$+18$ 10 53.17&$ 4.6060\p0.0146$&$-78.011\p0.019$&$-29.825\p0.016$& 12.388&$ 1.013$&no                  \\
M53 &2 &13 12 47.541&$+18$ 10 23.15&$ 1.1056\p0.0206$&$-17.135\p0.026$&$ -4.531\p0.024$& 14.395&$ 0.903$&no                  \\
    &3 &13 12 45.206&$+18$ 11 57.44&$ 0.0211\p0.0269$&$ -0.165\p0.036$&$ -1.351\p0.033$& 14.850&$ 0.192$&yes  & AHB   &  \rn \\
    &4 &13 12 47.406&$+18$ 07 10.77&$ 0.0519\p0.0505$&$ -0.401\p0.065$&$ -1.422\p0.062$& 15.701&$ 0.297$&yes  & AHB          \\
    &5 &13 13 09.478&$+18$ 10 58.70&$ 0.9811\p0.0241$&$-15.057\p0.032$&$-24.435\p0.031$& 14.741&$ 0.770$&no                  \\
\noalign{\smallskip}
    &6 &13 13 11.459&$+18$ 08 09.47&$ 0.0533\p0.0361$&$ -0.080\p0.049$&$ -1.297\p0.047$& 15.734&$ 1.136$&yes  & RGB          \\
    &7 &13 12 48.140&$+18$ 09 45.71&$-0.3570\p0.0836$&$ -0.588\p0.100$&$ -0.379\p0.100$& 15.129&$ 1.133$&yes? & AGB   &  \rn \\
    &8 &13 13 02.930&$+18$ 11 06.60&$-0.2584\p0.0964$&$ -0.627\p0.124$&$ -0.644\p0.125$& 14.015&$ 1.332$&yes  & AGB   &  \rn \\
    &9 &13 13 03.347&$+18$ 11 43.51&$ 0.0281\p0.0225$&$ -0.096\p0.030$&$ -1.392\p0.029$& 14.613&$ 1.292$&yes  & RGB   &  \rn \\
    &10&13 13 11.209&$+18$ 13 08.27&$ 1.0669\p0.0238$&$  0.680\p0.032$&$-21.557\p0.031$& 14.764&$ 0.808$&no                  \\
\noalign{\smallskip}
    &11&13 12 29.048&$+18$ 07 16.08&$ 1.3456\p0.0185$&$ -8.315\p0.024$&$ 10.187\p0.025$& 14.137&$ 0.776$&no                  \\
    &12&13 13 22.626&$+18$ 10 09.39&$ 1.0159\p0.0362$&$  2.877\p0.048$&$-10.336\p0.048$& 14.872&$ 0.842$&no                  \\
    &13&13 12 57.021&$+18$ 10 34.92&$-0.1280\p0.0818$&$ -0.573\p0.105$&$ -1.402\p0.100$& 14.309&$ 1.177$&yes  & AGB          \\ 
    &14&13 12 54.119&$+18$ 09 56.99&$ 0.3272\p0.2388$&$  1.954\p0.319$&$ -4.473\p0.376$& 13.968&$ 1.111$&no?  &       &  \rn \\
    &15&13 12 58.238&$+18$ 08 36.84&$ 2.7637\p0.2488$&$  3.468\p0.323$&$ -0.208\p0.313$& 15.881&$ 0.802$&no                  \\
\noalign{\smallskip}
    &16&13 12 54.235&$+18$ 11 47.54&$-0.0415\p0.0832$&$ -0.122\p0.111$&$ -1.451\p0.101$& 16.995&$ 0.197$&yes  & HB           \\
\noalign{\smallskip}
5053&1 &13 16 39.651&$+17$ 41 39.43&$ 1.5468\p0.0237$&$ -7.021\p0.025$&$  4.437\p0.026$& 14.461&$ 0.923$&no                  \\
    &2 &13 16 08.506&$+17$ 41 43.35&$ 1.1008\p0.0251$&$-13.801\p0.029$&$ -8.455\p0.027$& 14.411&$ 0.821$&no                  \\
\noalign{\smallskip}
5272/&1 &13 42 16.750&$+28$ 26 00.61&$ 0.0387\p0.0413$&$ -0.241\p0.050$&$ -2.716\p0.028$& 14.947&$-0.442$&yes  & vPAGB &  \rn \\ 
M3  &2 &13 42 10.767&$+28$ 19 05.32&$ 1.9252\p0.0153$&$-10.787\p0.018$&$  5.088\p0.010$& 13.343&$ 0.859$&no                  \\
    &3 &13 42 23.099&$+28$ 25 00.53&$ 0.0417\p0.0415$&$ -0.115\p0.045$&$ -2.521\p0.025$& 15.621&$ 0.637$&yes  & HB    &  \rn \\ 
    &4 &13 42 17.042&$+28$ 23 02.81&$ 0.0561\p0.1787$&$ -0.243\p0.192$&$ -1.689\p0.118$& 15.616&$ 0.669$&yes  & HB    &  \rn \\ 
    &5 &13 42 18.924&$+28$ 19 34.11&$ 0.0615\p0.0368$&$ -0.291\p0.042$&$ -2.495\p0.024$& 15.657&$ 0.551$&yes  & HB    &  \rn \\ 
\noalign{\smallskip}
    &6 &13 41 51.504&$+28$ 17 44.54&$ 0.7638\p0.0203$&$-13.185\p0.022$&$ -4.738\p0.012$& 13.931&$ 0.813$&no                  \\
\noalign{\smallskip}
5466&1 &14 05 10.675&$+28$ 28 54.23&$ 0.6579\p0.0257$&$  7.710\p0.027$&$-13.591\p0.022$& 13.861&$ 0.592$&no                  \\  
    &2 &14 05 15.075&$+28$ 26 52.20&$ 1.4961\p0.0184$&$ 19.140\p0.019$&$-17.784\p0.016$& 13.987&$ 0.874$&no                  \\
    &3 &14 05 16.967&$+28$ 25 46.55&$ 1.1790\p0.0158$&$ -4.106\p0.017$&$ -6.210\p0.014$& 13.867&$ 0.719$&no                  \\
    &4 &14 05 38.767&$+28$ 33 00.05&$ 0.1491\p0.0185$&$  0.447\p0.020$&$-11.631\p0.017$& 14.060&$ 0.818$&no                  \\
    &5 &14 05 41.566&$+28$ 34 37.32&$-0.1447\p0.0802$&$ -5.302\p0.093$&$ -0.753\p0.075$& 17.410&$-0.144$&yes  & HB    &  \rn \\
\noalign{\smallskip}
    &6 &14 05 29.091&$+28$ 32 47.23&$ 0.0124\p0.0176$&$ -5.360\p0.019$&$ -0.868\p0.016$& 14.188&$ 1.139$&yes  & AGB   &  \rn \\
\noalign{\smallskip}
5634&1 &14 29 36.219&$-05$ 58 16.76&$ 0.2767\p0.0423$&$  0.074\p0.053$&$-26.117\p0.042$& 15.650&$ 0.762$&no                  \\
    &2 &14 29 33.211&$-05$ 57 45.36&$ 0.5259\p0.0181$&$-19.371\p0.023$&$-21.113\p0.019$& 14.080&$ 1.030$&no                  \\ 
    &3 &14 29 29.487&$-05$ 58 44.27&$ 0.9716\p0.0202$&$ -3.394\p0.026$&$ -9.159\p0.021$& 13.988&$ 0.828$&no                  \\
\noalign{\smallskip}
5897&1 &15 17 17.592&$-21$ 03 18.75&$ 0.2510\p0.0173$&$ -4.685\p0.017$&$ -7.882\p0.014$& 13.330&$ 0.961$&no                  \\
    &2 &15 17 30.443&$-21$ 00 10.37&$ 0.1025\p0.0178$&$ -5.433\p0.017$&$ -3.241\p0.016$& 12.672&$ 1.378$&yes  & rPAGB &  \rn \\
\noalign{\smallskip} 
5904/&1 &15 18 34.101&$+02$ 04 59.76&$ \dots$&$  \dots$&$\dots$& $\dots$&$ \dots$&yes  & vPAGB      &  \rn \\
M5  &2 &15 18 35.690&$+02$ 03 46.11&$-0.1568\p0.0394$&$  4.195\p0.049$&$ -9.910\p0.038$& 11.562&$ 1.705$&no?  &       &  \rn \\
    &3 &15 18 17.349&$+02$ 02 29.55&$ 1.4686\p0.0485$&$ -7.348\p0.054$&$ -2.279\p0.054$& 13.317&$ 0.874$&no                  \\
    &4 &15 18 48.160&$+02$ 02 16.81&$ 1.5443\p0.0181$&$-23.625\p0.019$&$ 17.194\p0.021$& 13.422&$ 0.846$&no                  \\
    &5 &15 18 45.927&$+02$ 04 14.90&$ 1.0363\p0.0226$&$-17.570\p0.024$&$ -1.869\p0.024$& 13.776&$ 0.811$&no                  \\   
\noalign{\smallskip}
    &6 &15 18 35.026&$+02$ 08 40.09&$ 0.0479\p0.0338$&$  4.147\p0.031$&$ -9.885\p0.031$& 15.123&$ 0.617$&yes  & HB    &  \rn \\
    &7 &15 18 31.070&$+02$ 02 42.76&$-0.1631\p0.1482$&$  3.869\p0.132$&$ -9.659\p0.129$& 14.574&$ 0.879$&yes  & HB    &  \rn \\
\noalign{\smallskip}
6093/&1 &16 17 04.351&$-22$ 59 11.72&$ 0.2111\p0.0280$&$ -2.942\p0.034$&$ -5.821\p0.029$& 13.660&$ 1.573$&no?  &       &  \rn \\
M80 &2 &16 17 15.936&$-22$ 58 28.32&$ 1.7916\p0.0194$&$ -3.290\p0.027$&$ -1.695\p0.021$& 13.826&$ 0.982$&no                  \\ 
    &3 &16 17 01.555&$-22$ 58 30.68&$       \dots   $&$       \dots  $&$       \dots  $& 15.216&$ 1.313$&yes? & CV?   &  \rn \\  
    &4 &16 17 11.479&$-23$ 01 57.38&$ 1.8891\p0.1118$&$-13.539\p0.131$&$-13.728\p0.100$& 13.842&$ 1.131$&no                  \\
    &5 &16 17 08.152&$-22$ 56 32.05&$ 2.1391\p0.0187$&$-12.222\p0.024$&$-14.064\p0.020$& 12.124&$ 0.898$&no                  \\
\noalign{\smallskip}
6121/&1 &16 23 32.449&$-26$ 26 45.86&$ 5.0588\p0.0206$&$  5.387\p0.026$&$ -8.654\p0.019$& 11.182&$ 1.128$&no                  \\
M4  &2 &16 23 35.031&$-26$ 25 36.49&$ 0.5575\p0.0189$&$-13.069\p0.024$&$-18.276\p0.018$& 13.183&$ 1.087$&yes  & HB    &  \rn \\
    &3 &16 23 39.404&$-26$ 34 54.53&$ 4.7764\p0.0205$&$ 17.849\p0.025$&$-15.673\p0.018$& 10.103&$ 1.189$&no                  \\
    &4 &16 23 29.185&$-26$ 28 54.38&$ 0.5858\p0.0205$&$-12.286\p0.025$&$-18.590\p0.018$& 13.161&$ 1.095$&yes  & HB    &  \rn \\
    &5 &16 23 36.469&$-26$ 30 44.33&$ 0.6220\p0.0239$&$-11.733\p0.029$&$-20.013\p0.022$& 12.894&$ 0.992$&yes  & HB    &  \rn \\
\noalign{\smallskip}
    &6 &16 23 47.824&$-26$ 32 05.99&$ 0.5006\p0.0165$&$-12.794\p0.022$&$-20.022\p0.016$& 13.395&$ 0.605$&yes  & HB    &  \rn \\
    &7 &16 23 48.797&$-26$ 35 09.17&$ 0.5526\p0.0467$&$-11.702\p0.059$&$-18.739\p0.036$& 11.026&$ 1.888$&yes  & RGB   &  \rn \\
    &8 &16 23 50.103&$-26$ 36 09.55&$ 0.5132\p0.0188$&$-12.461\p0.022$&$-18.989\p0.017$& 12.894&$ 0.731$&yes  & HB           \\   
    &9 &16 23 53.591&$-26$ 30 05.35&$ 0.5757\p0.0221$&$-12.615\p0.028$&$-19.512\p0.021$& 12.946&$ 1.016$&yes  & HB    &  \rn \\
\noalign{\smallskip}
6205/&1 &16 41 33.666&$+36$ 26 07.78&$ 0.0775\p0.0298$&$ -3.203\p0.031$&$ -2.760\p0.033$& 13.123&$-0.255$&yes  & vPAGB &  \rn \\
M13 &2 &16 41 34.759&$+36$ 29 13.89&$ 2.0187\p0.4059$&$ -3.190\p0.411$&$ -3.180\p0.464$& 14.270&$ 0.267$&yes? & AHB   &  \rn \\
    &3 &16 41 52.090&$+36$ 26 28.90&$ 0.0934\p0.0164$&$ -3.137\p0.015$&$ -2.712\p0.018$& 14.089&$ 0.182$&yes  & AHB   &  \rn \\
    &4 &16 41 36.437&$+36$ 30 51.67&$ 0.1158\p0.0174$&$ -3.229\p0.017$&$ -2.572\p0.021$& 14.055&$ 0.206$&yes  & AHB   &  \rn \\
    &5 &16 41 19.580&$+36$ 21 15.62&$ 0.2542\p0.0146$&$ -4.672\p0.015$&$ -1.707\p0.019$& 11.380&$ 1.442$&no                  \\
\noalign{\smallskip}
    &6 &16 41 43.018&$+36$ 28 42.20&$ 0.0894\p0.0228$&$ -3.397\p0.021$&$ -2.414\p0.027$& 14.646&$ 0.014$&yes  & AHB          \\ 
    &7 &16 41 08.024&$+36$ 30 05.38&$ 0.0878\p0.0250$&$ -3.287\p0.025$&$ -2.469\p0.034$& 15.361&$-0.185$&yes  & AHB          \\  
\noalign{\smallskip}
6218/&1 &16 47 18.547&$-02$ 00 10.86&$ 1.8183\p0.0132$&$-57.328\p0.014$&$-89.440\p0.012$& 12.326&$ 1.005$&no                  \\
M12 &2 &16 47 00.790&$-01$ 58 28.25&$ 0.3748\p0.0137$&$ -9.177\p0.015$&$-21.670\p0.013$& 12.587&$ 1.294$&no                  \\
    &3 &16 47 12.136&$-01$ 54 06.92&$ 1.2808\p0.0132$&$ -2.300\p0.015$&$  8.715\p0.012$& 11.926&$ 1.300$&no                  \\
    &4 &16 47 08.603&$-02$ 01 37.07&$ 0.4723\p0.0123$&$ -1.061\p0.014$&$ -4.560\p0.011$& 12.479&$ 1.338$&no                  \\
    &5 &16 47 45.287&$-01$ 55 03.50&$ 0.7490\p0.0136$&$ -7.806\p0.015$&$ -6.641\p0.012$& 13.183&$ 0.915$&no                  \\ 
\noalign{\smallskip}
    &6 &16 47 27.241&$-01$ 52 19.49&$ 1.2837\p0.0151$&$  6.753\p0.019$&$ -8.277\p0.015$& 13.590&$ 0.921$&no                  \\
    &7 &16 47 18.056&$-01$ 58 17.40&$ 0.1382\p0.0135$&$ -0.358\p0.015$&$ -6.911\p0.013$& 12.690&$ 1.401$&yes  & AGB  & \rn         \\
    &8 &16 47 25.974&$-02$ 01 03.39&$ 0.1799\p0.0119$&$ -0.187\p0.013$&$ -6.882\p0.010$& 12.136&$ 1.435$&yes  & AGB  & \rn        \\
    &9 &16 47 39.628&$-02$ 01 25.49&$ 1.7309\p0.0176$&$ 15.382\p0.020$&$ -1.958\p0.016$& 13.089&$ 0.876$&no                  \\
\noalign{\smallskip}
6254/&1 &16 57 09.253&$-04$ 04 24.40&$ 0.1131\p0.0230$&$ -4.844\p0.025$&$ -6.795\p0.019$& 13.485&$ 0.016$&yes  & bPAGB &  \rn \\
M10 &2 &16 57 09.836&$-04$ 04 28.50&$ 0.1663\p0.0205$&$ -4.342\p0.023$&$ -6.248\p0.017$& 13.729&$ 0.354$&yes  & AHB          \\
    &3 &16 57 18.697&$-04$ 09 06.82&$ 0.9588\p0.0262$&$  0.859\p0.027$&$ -5.637\p0.020$& 11.152&$ 1.434$&no                  \\
    &4 &16 57 10.387&$-04$ 04 01.74&$ 0.1697\p0.0170$&$ -4.934\p0.018$&$ -6.616\p0.014$& 13.026&$ 1.525$&yes  & RGB          \\
    &5 &16 57 05.571&$-04$ 06 52.31&$ 0.2817\p0.0335$&$ -4.618\p0.035$&$ -6.585\p0.029$& 13.998&$ 0.426$&yes  & AHB          \\
\noalign{\smallskip}
    &6 &16 57 06.332&$-04$ 07 17.03&$ 0.1471\p0.0734$&$ -4.279\p0.082$&$ -6.317\p0.069$& 13.940&$ 0.741$&yes  & AHB          \\
    &7 &16 57 12.231&$-04$ 11 26.97&$ 1.7263\p0.0232$&$ -0.705\p0.024$&$  2.203\p0.019$& 11.398&$ 1.392$&no                  \\
    &8 &16 56 57.629&$-04$ 07 16.03&$ 0.2181\p0.0229$&$ -4.786\p0.024$&$ -6.737\p0.018$& 14.248&$ 0.376$&yes  & AHB          \\
\noalign{\smallskip}
6273/&1 &17 02 41.532&$-26$ 15 16.42&$ 0.9908\p0.0165$&$ -1.240\p0.018$&$ -2.328\p0.012$& 12.083&$ 0.526$&no                  \\
M19 &2 &17 02 39.154&$-26$ 15 29.36&$ 0.1582\p0.0236$&$ -2.990\p0.025$&$  1.454\p0.018$& 13.207&$ 0.518$&yes  & bPAGB &  \rn \\   
    &3 &17 02 38.139&$-26$ 15 11.75&$ 0.3344\p0.0409$&$ -3.034\p0.048$&$  1.946\p0.033$& 13.342&$ 1.404$&yes? & AHB   &  \rn \\
    &4 &17 02 35.185&$-26$ 15 24.13&$ 0.1177\p0.0157$&$ -2.878\p0.019$&$  1.146\p0.012$& 12.204&$ 1.238$&yes  & yPAGB &  \rn \\
    &5 &17 02 24.171&$-26$ 14 40.71&$ 1.8255\p0.0176$&$ -4.142\p0.019$&$ -2.644\p0.013$& 12.718&$ 0.931$&no                  \\
\noalign{\smallskip}
    &6 &17 02 26.447&$-26$ 16 08.28&$ 1.2297\p0.4424$&$ -3.274\p0.508$&$-10.004\p0.353$& 13.570&$ 1.285$&no                  \\
\noalign{\smallskip}
6333/&1 &17 19 15.461&$-18$ 31 10.71&$ 0.0697\p0.0183$&$ -2.326\p0.020$&$ -3.390\p0.014$& 12.836&$ 0.561$&yes? & bPAGB &  \rn \\
M9  &2 &17 19 05.776&$-18$ 32 53.32&$ 0.3581\p0.0150$&$  6.749\p0.018$&$-14.170\p0.012$& 13.187&$ 0.569$&no                  \\         
    &3 &17 19 03.581&$-18$ 30 43.67&$ 2.5533\p0.0162$&$  0.193\p0.019$&$ -0.128\p0.012$& 13.308&$ 1.147$&no                  \\
    &4 &17 19 05.924&$-18$ 31 33.16&$ 0.3946\p0.1833$&$ -2.197\p0.212$&$ -4.114\p0.137$& 14.480&$ 0.787$&yes? & AHB   &  \rn \\
\noalign{\smallskip}
6341/&1 &17 16 56.770&$+43$ 05 36.82&$ 2.0357\p0.0127$&$  5.480\p0.015$&$ -2.531\p0.015$& 11.670&$ 0.576$&no                  \\
M92 &2 &17 17 03.850&$+43$ 05 50.45&$ 4.9922\p0.0386$&$-56.771\p0.046$&$-63.174\p0.047$& 12.679&$ 1.085$&no                  \\
    &3 &17 16 58.376&$+43$ 01 18.04&$ 1.0549\p0.0130$&$ -8.175\p0.016$&$  0.250\p0.016$& 12.953&$ 0.775$&no                  \\
    &4 &17 16 29.018&$+43$ 09 43.80&$ 0.2461\p0.0105$&$ -4.061\p0.013$&$ -1.535\p0.012$& 13.328&$ 0.943$&no          &    & \rn    \\
\noalign{\smallskip}
6356&1 &17 23 32.455&$-17$ 49 38.41&$ 1.1879\p0.0297$&$  2.353\p0.030$&$ -6.151\p0.020$& 14.380&$ 1.101$&no                  \\
    &2 &17 23 29.547&$-17$ 49 10.62&$ 0.0568\p0.0294$&$  1.815\p0.036$&$ -3.089\p0.021$& 14.265&$ 1.482$&yes? & AHB   &  \rn \\ 
    &3 &17 23 40.827&$-17$ 49 24.47&$ 0.4388\p0.1070$&$ -4.023\p0.099$&$  1.345\p0.074$& 14.820&$ 1.295$&no                  \\
\noalign{\smallskip} 
6402/&1 &17 37 33.178&$-03$ 14 51.68&$ 0.0748\p0.0368$&$ -3.803\p0.034$&$ -5.178\p0.026$& 14.538&$ 0.863$&yes  & bPAGB &  \rn \\
M14 &2 &17 37 33.434&$-03$ 16 09.89&$ 1.3874\p0.0193$&$ -4.388\p0.019$&$  5.385\p0.014$& 13.217&$ 1.356$&no                  \\ 
    &3 &17 37 44.596&$-03$ 11 50.37&$ 0.5030\p0.0229$&$  4.054\p0.022$&$ -1.104\p0.017$& 13.696&$ 1.000$&no                  \\
    &4 &17 37 48.410&$-03$ 15 47.78&$ 2.2708\p0.0663$&$  0.130\p0.063$&$-10.114\p0.050$& 13.446&$ 1.482$&no                  \\
    &5 &17 37 30.986&$-03$ 15 03.77&$ 0.1841\p0.0449$&$ -3.501\p0.043$&$ -5.080\p0.033$& 15.257&$ 0.842$&yes  & AHB   &  \rn \\
\noalign{\smallskip}
    &6 &17 37 33.749&$-03$ 15 42.09&$-0.0981\p0.0531$&$ -3.659\p0.051$&$ -4.962\p0.039$& 14.642&$ 1.128$&yes  & AHB          \\
    &7 &17 37 31.877&$-03$ 14 52.47&$-0.0424\p0.0385$&$ -3.687\p0.037$&$ -5.033\p0.029$& 15.353&$ 0.966$&yes  & AHB          \\ 
    &8 &17 37 21.521&$-03$ 15 18.01&$ 1.1383\p0.1176$&$  4.360\p0.094$&$  0.377\p0.074$& 14.727&$ 1.345$&no                  \\ 
    &9 &17 37 27.238&$-03$ 12 56.63&$ 1.3406\p0.0259$&$  1.469\p0.024$&$ 10.036\p0.019$& 14.759&$ 1.510$&no                  \\
    &10&17 37 35.767&$-03$ 16 29.98&$ 0.1324\p0.0385$&$ -3.612\p0.035$&$ -4.943\p0.028$& 15.569&$ 0.916$&yes  & AHB          \\   
\noalign{\smallskip}
    &11&17 37 48.235&$-03$ 15 52.97&$ 2.0882\p0.0188$&$  0.019\p0.018$&$-10.942\p0.014$& 13.862&$ 1.458$&no                  \\
    &12&17 37 42.232&$-03$ 14 59.35&$ 0.1898\p0.0961$&$ -3.961\p0.088$&$ -5.138\p0.071$& 15.622&$ 0.895$&yes  & AHB          \\
    &13&17 37 26.563&$-03$ 17 15.06&$ 0.5451\p0.0140$&$ -2.361\p0.013$&$ -1.307\p0.010$& 12.538&$ 1.842$&no                  \\
    &14&17 37 33.474&$-03$ 15 27.77&$-0.0213\p0.0339$&$ -3.825\p0.033$&$ -5.680\p0.026$& 14.980&$ 1.689$&yes  & AHB   &  \rn \\
\noalign{\smallskip} 
6626/&1 &18 24 34.602&$-24$ 53 21.19&$ 0.5888\p0.0213$&$  0.742\p0.034$&$ -0.457\p0.024$& 12.928&$ 0.765$&no                  \\
M28 &2 &18 24 28.326&$-24$ 54 24.28&$ 0.7298\p0.0293$&$  1.540\p0.039$&$ -0.620\p0.028$& 12.760&$ 0.599$&no                  \\
    &3 &18 24 29.356&$-24$ 50 47.10&$ 0.5752\p0.0282$&$  0.703\p0.042$&$  0.210\p0.030$& 13.733&$ 0.815$&no                  \\
    &4 &18 24 31.597&$-24$ 49 16.99&$ 1.1453\p0.0205$&$  0.719\p0.023$&$ -3.168\p0.017$& 12.554&$ 0.762$&no                  \\ 
    &5 &18 24 35.838&$-24$ 53 15.87&$ 0.1444\p0.0242$&$  0.137\p0.030$&$ -8.671\p0.021$& 11.610&$ 1.755$&yes  & rPAGB &  \rn \\
\noalign{\smallskip}
6656/&1 &18 36 13.295&$-23$ 52 46.13&$ 4.3374\p0.0167$&$ 12.512\p0.018$&$ 11.282\p0.013$& 10.969&$ 1.043$&no                  \\   
M22 &2 &18 36 40.101&$-23$ 52 51.13&$ 3.4277\p0.0179$&$ 15.181\p0.017$&$ -4.332\p0.013$& 11.812&$ 1.011$&no                  \\
    &3 &18 36 49.468&$-23$ 55 48.96&$ 1.0871\p0.0164$&$ -0.813\p0.016$&$ -4.461\p0.013$& 11.863&$ 0.983$&no                  \\
    &4 &18 36 35.319&$-23$ 48 20.50&$ 1.2092\p0.0432$&$  2.420\p0.038$&$ -4.826\p0.028$& 11.576&$ 0.697$&no                  \\
    &5 &18 36 10.796&$-23$ 49 57.14&$ 0.3008\p0.0180$&$  9.250\p0.019$&$ -5.808\p0.014$& 12.519&$ 0.668$&yes  & AHB   &      \\
\noalign{\smallskip}
    &6 &18 36 30.544&$-23$ 53 58.10&$ 0.2797\p0.0181$&$ 10.187\p0.020$&$ -5.710\p0.017$& 10.970&$ 1.805$&yes  & AGB          \\
    &7 &18 36 24.181&$-23$ 52 57.09&$ 0.2833\p0.0185$&$ 10.493\p0.021$&$ -6.293\p0.014$& 10.782&$ 1.975$&yes  & RGB          \\
    &8 &18 35 50.312&$-23$ 58 25.91&$ 1.0202\p0.0437$&$  0.880\p0.047$&$ -4.359\p0.038$& 10.999&$ 0.578$&no                  \\
    &9 &18 36 49.548&$-24$ 01 16.01&$ 0.2440\p0.1104$&$ -2.543\p0.127$&$ -6.754\p0.116$& 11.898&$ 0.950$&no                  \\
    &10&18 35 52.789&$-23$ 53 34.20&$ 1.0688\p0.0215$&$  0.383\p0.022$&$ -3.565\p0.016$& 10.210&$ 1.730$&no                  \\
\noalign{\smallskip}
    &11&18 36 33.013&$-23$ 54 35.71&$ 1.4884\p0.0158$&$  5.226\p0.018$&$-14.614\p0.015$& 10.999&$ 1.438$&no                  \\
    &12&18 36 15.426&$-23$ 46 17.40&$ 1.0804\p0.0182$&$  1.260\p0.021$&$  1.393\p0.017$& 12.369&$ 0.501$&no                  \\
    &13&18 36 10.475&$-23$ 46 27.45&$ 1.5638\p0.0968$&$  3.307\p0.099$&$ -2.470\p0.080$& 10.501&$ 0.537$&no                  \\
    &14&18 35 54.081&$-23$ 45 31.73&$ 0.2634\p0.0188$&$  9.807\p0.021$&$ -4.879\p0.016$& 10.931&$ 1.750$&yes  & AGB  & \rn         \\
    &15&18 35 43.009&$-23$ 48 32.99&$ 5.2690\p0.0178$&$-46.562\p0.022$&$-77.378\p0.016$& 10.792&$ 0.985$&no                  \\
\noalign{\smallskip}
    &16&18 36 29.196&$-23$ 49 53.24&$ 0.3072\p0.0233$&$  4.747\p0.022$&$ -0.703\p0.016$& 13.520&$ 0.356$&no                  \\
    &17&18 36 53.525&$-23$ 51 29.50&$ 1.2703\p0.0167$&$  0.881\p0.017$&$  0.937\p0.013$&  9.739&$ 1.431$&no                  \\
    &18&18 36 28.974&$-24$ 04 04.05&$ 1.5693\p0.0174$&$  4.618\p0.020$&$ -6.379\p0.016$& 12.822&$ 0.949$&no                  \\
\noalign{\smallskip}
6712&1 &18 53 05.791&$-08$ 42 37.85&$ 0.0890\p0.0156$&$  3.275\p0.018$&$ -4.515\p0.014$& 13.135&$ 0.556$&yes  & bPAGB &  \rn \\
    &2 &18 53 07.295&$-08$ 43 09.29&$ 0.3587\p0.0250$&$  0.766\p0.027$&$  0.328\p0.021$& 13.870&$ 0.797$&no                  \\
    &3 &18 53 01.434&$-08$ 42 33.89&$ 1.2943\p0.0212$&$  6.463\p0.025$&$ -0.903\p0.022$& 14.109&$ 0.986$&no                  \\
    &4 &18 53 08.772&$-08$ 41 56.66&$ 0.1354\p0.0196$&$  3.367\p0.021$&$ -4.476\p0.017$& 12.025&$ 2.443$&yes  & AGB   &  \rn \\
\noalign{\smallskip}
6779/&1 &19 16 36.583&$+30$ 11 17.37&$ 0.3612\p0.1729$&$ -6.241\p0.160$&$-18.053\p0.183$& 14.423&$ 0.969$&no                  \\
M56 &2 &19 16 41.278&$+30$ 12 48.55&$ 0.0486\p0.0237$&$ -1.946\p0.021$&$  1.598\p0.025$& 15.099&$-0.038$&yes  & vPAGB &  \rn \\
\noalign{\smallskip}
6934&1 &20 34 07.304&$+07$ 23 47.01&$ 0.9455\p0.0213$&$ 12.508\p0.023$&$ -2.894\p0.017$& 14.528&$ 0.874$&no                  \\
\noalign{\smallskip}
7078/&1 &21 29 58.194&$+12$ 11 42.55&$ 0.0825\p0.0435$&$ -0.631\p0.047$&$ -3.736\p0.039$& 15.065&$ 0.053$&yes  & AHB   &  \rn \\
M15 &2 &21 30 07.438&$+12$ 11 07.73&$ 0.0962\p0.0217$&$ -0.568\p0.022$&$ -3.859\p0.017$& 14.313&$ 0.137$&yes  & AHB          \\
    &3 &21 29 44.092&$+12$ 09 16.73&$ 1.1896\p0.0149$&$ -2.117\p0.014$&$ -3.762\p0.012$& 13.477&$ 0.910$&no                  \\
    &4 &21 30 00.789&$+12$ 09 43.28&$-0.5487\p0.2111$&$ -0.971\p0.219$&$ -2.189\p0.192$& 13.820&$ 1.160$&yes  & AGB   &  \rn \\  
    &5 &21 29 55.072&$+12$ 10 14.49&$-0.1125\p0.0343$&$ -0.398\p0.036$&$ -3.492\p0.033$& 13.119&$ 1.373$&yes  & AGB   &  \rn \\
\noalign{\smallskip}
    &6 &21 29 56.179&$+12$ 10 17.93&$ 0.3868\p0.0513$&$ -0.405\p0.052$&$ -4.058\p0.045$& 12.352&$ 1.511$&no?  &       &  \rn \\
    &7 &21 30 00.981&$+12$ 08 41.83&$-0.4838\p0.1057$&$ -0.638\p0.109$&$ -2.803\p0.091$& 15.194&$ 0.230$&yes? & AHB   &  \rn \\
\noalign{\smallskip} 
7089/&1 &21 33 34.128&$-00$ 53 37.99&$ 0.0232\p0.0324$&$  3.278\p0.034$&$ -2.200\p0.021$& 14.570&$ 0.139$&yes  & AHB          \\
M2  &2 &21 33 29.872&$-00$ 45 51.60&$-0.0170\p0.0773$&$  3.276\p0.091$&$ -2.724\p0.065$& 15.113&$ 0.162$&yes  & AHB          \\
\enddata
\tablenotetext{a}{Indication of cluster membership, based on parallax and proper motion. A few uncertain cases are labeled with question marks.}  
\tablenotetext{b}{Evolutionary classification scheme is explained in text (\S\ref{sec:results}) and Figure~2 caption; based on position in \Gaia\/ color-magnitude diagrams. Abbreviations are: AGB (asymptotic-giant branch); RGB (red-giant branch); HB (horizontal branch); rPAGB, yPAGB, bPAGB, and vPAGB (red, yellow, blue, and violet post-AGB stars); AHB (above horizontal branch); and CV (cataclysmic variable).}
\tablenotetext{c}{{  Remarks:} 
[\nn] 4147 ZNG~2: \HST\/ images show object to be blend of 5 cluster members of
similar brightness; brightest one listed. 
[\nn] 4590 ZNG~2: At tip of AGB\slash RGB; see \citet{Schaeuble2015} and references therein.  
[\nn] 5024 ZNG~3: considered a non-member by \citet{Boberg2016}. 
[\nn] 5024 ZNG~7: Close nearly equal pair, separation 0".9. Astrometry has large uncertainties, but both components
consistent with membership. Brighter component listed. Non-member according to \citet{Boberg2016}.
[\nn] 5024 ZNG~8: Astrometry has large uncertainties. Member according to \citet{Boberg2016}.
[\nn] 5024 ZNG~9: Member according to \citet{Boberg2016}. Blended with a fainter AGB star $2\farcs0$ away. 
[\nn] 5024 ZNG~14: Identification uncertain. Astrometry has large uncertainties, but membership unlikely. 
[\nn] 5272 ZNG~1: M3 von~Zeipel 1128.
[\nn] 5272 ZNG~3: RR~Lyr variable V10.
[\nn] 5272 ZNG~4: Blend of 3 stars; bluest one listed. RR~Lyr variable V165.
[\nn] 5272 ZNG~5: RR~Lyr variable V90.
[\nn] 5466 ZNG~5: Extreme HB star; see \citet{Geier2019}.
[\nn] 5466 ZNG~6: AGB star; see \citet{Shetrone2010}.
[\nn] 5897 ZNG~2: Luminous red post-AGB star. \Gaia\/ RV consistent with membership.
[\nn] 5904 ZNG~1: Known hot vPAGB star (e.g., \citealt{Dixon2004}), not contained in EDR3 because of neighboring ($0\farcs52$) bright red giant; position obtained from \HST\/ images.
[\nn] 5904 ZNG~2: Proper motion consistent with membership, parallax discordant; AGB or RGB tip star if member.
[\nn] 5904 ZNG~6: RR~Lyr variable V41.
[\nn] 5904 ZNG~7: RR~Lyr variable V25.
[\nn] 6093 ZNG~1: Proper motion consistent with membership, parallax discordant; RGB star if member. 
[\nn] 6093 ZNG~3: EDR3 does not give a parallax or proper motion for this source; there is a bright blue object at this position in \HST\/ images, but the \Gaia\/ $BP-RP$ color is red. Position is close to that of the X-ray source CX3 \citep{Heinke2003}.  
[\nn] 6121 ZNG~2: RR~Lyr variable V19.
[\nn] 6121 ZNG~4: RR~Lyr variable V10.
[\nn] 6121 ZNG~5: This object is a blend of two RR~Lyr variables, separated by $1\farcs6$ \citep{Clementini2019}. Brighter component is listed in table.
[\nn] 6121 ZNG~6: Candidate RR~Lyr, V58, but non-variable according to \citet{Safonova2016}. 
[\nn] 6121 ZNG~7: Very bright star at tip of RGB\null. \Gaia\/ RV consistent with membership.
[\nn] 6121 ZNG~9: RR~Lyr variable V28.
[\nn] 6205 ZNG~1: Barnard 29 (see M+19).
[\nn] 6205 ZNG~2: Proper motion consistent with membership, but large uncertainties and discordant parallax; suspected low-amplitude variable \citep{Deras2019}.
[\nn] 6205 ZNG~3: Variable star V51, uncertain type \citep{Servillat2011}. 
[\nn] 6205 ZNG~4: Spectroscopic analysis by \citet{Ambika2004}.
[\nn] 6218 ZNG~7: \Gaia\/ RV consistent with membership.
[\nn] 6218 ZNG~8: \Gaia\/ RV consistent with membership.
[\nn] 6254 ZNG~1: Blue PAGB star (see M+19).
[\nn] 6273 ZNG~2: Blue PAGB star (see B+21).
[\nn] 6273 ZNG~3: Type~II Cepheid, V1 \citep{Clement1978}; proper motion consistent with membership, parallax discordant.
[\nn] 6273 ZNG~4: Yellow PAGB star (see B+21).
[\nn] 6333 ZNG~1: Proper motion consistent with membership; parallax slightly discrepant. On boundary between yellow and blue PAGB stars. 
[\nn] 6333 ZNG~4: Astrometry consistent with membership, but uncertainties are large.
[\nn] 6341 ZNG~4: Proper motion consistent with membership, but parallax is discordant. Considered a member by \citet{Zinn1974}, but not by \citet{Roederer2011}.
[\nn] 6356 ZNG~2: Proper motion marginally consistent with membership.
[\nn] 6402 ZNG~1: Blue PAGB star (star 33 in \citealt{Schiavon2012}).
[\nn] 6402 ZNG~5: Above HB (star 202 in \citealt{Schiavon2012}).
[\nn] 6402 ZNG~14: Type II Cepheid, V167 \citep{Contreras2018}.
[\nn] 6626 ZNG~5: RV~Tauri variable V2342 Sgr = V17.
[\nn] 6656 ZNG~14: \Gaia\/ RV consistent with membership.
[\nn] 6712 ZNG~1: Blue PAGB star (see M+19).
[\nn] 6712 ZNG~4: Luminous star at AGB tip; Mira variable AP~Sct = V2.
[\nn] 6779 ZNG~2: Violet PAGB star (see M+19).
[\nn] 7078 ZNG~1: Hot post-HB or violet PAGB star (see M+19).
[\nn] 7078 ZNG~4: Identification difficult and uncertain. Listed star is brighter of close $1\farcs4$ pair. 
[\nn] 7078 ZNG~5: Identification difficult and uncertain. Listed star is brighter of close $1\farcs4$ pair. 
[\nn] 7078 ZNG~6: Proper motion consistent with membership, but parallax is discordant. If it is a member, it is
a bright star near the AGB tip.
[\nn] 7078 ZNG~7: Astrometry has large uncertainties; \citet{Gebhardt1997} give a radial velocity consistent with
membership.
}
\countnotes  
\end{deluxetable*}

\end{document}